%% file: submit.tex
\documentclass[aps,prl,twocolumn,superscriptaddress]{revtex4-1}

\usepackage{color}
\usepackage{amsmath}    
\usepackage{enumitem}
\usepackage{amssymb}
\usepackage{graphicx}   
\usepackage{subfigure}
\usepackage{float}

\begin{document}


\title{Anomalous Lifshitz dimension in hierarchical networks of brain connectivity}

\author{Samaneh Esfandiary}
\author{Ali Safari}
\author{Jakob Renner}
\author{Paolo Moretti}
\affiliation{Institute of Materials Simulation, FAU Universit\"at Erlangen-N\"urnberg, Dr.-Mack-Stra{\ss}e 77, 90762 F\"urth, Germany}
\author{Miguel A. Mu\~noz}
\affiliation{Departamento de Electromagnetismo y F{\'i}sica de la Materia
  e Instituto Carlos I de F{\'i}sica Te{\'o}rica y
  Computacional. Universidad de Granada, E-18071 Granada, Spain}

\date{\today}

\begin{abstract}
The spectral dimension is a generalization of the Euclidean dimension and quantifies the propensity of a network to transmit and diffuse information. We show that, in hierarchical-modular network models of the brain, dynamics are anomalously slow and the spectral dimension is not defined. Inspired by Anderson localization in quantum systems, we relate the localization of neural activity -- essential to embed brain functionality -- to the network spectrum and to the existence of an anomalous ``Lifshitz dimension''.  In a broader context, our results help shedding light on the relationship between structure and function in biological information-processing complex networks.
 \end{abstract}

\maketitle

Understanding the interplay between dynamical processes and the architecture of the networks embedding them is a fundamental problem in diverse fields including material science, genetic regulation and neuroscience. Dynamical features and patterns of activity are often affected or controlled by key structural features of the underlying network, such as the degree distribution, degree correlations, modular organization, k-core structure, etc. \cite{Liggett,Barrat,Redner,Newman,Barabasi}. However, given that such features are usually not independent, a more systematic way to tackle the problem of the interplay between structure and dynamics relies on the use of spectral-graph characterizations of the network architecture \cite{Chung,Mieghem} and, importantly, the network dimension. Statistical mechanics teaches us that dynamical processes such as diffusion, vibrational excitations, and critical properties near second order phase transitions exhibit universal behavior, which depends crucially on the lattice (Euclidean) dimension \cite{Alexander1982,Liggett,Barrat,Redner,Binney}. The case of heterogeneous networks is more complex, since multiple and diverse generalizations of the concept of dimension have been proposed \cite{Shanker1,Shanker2,Havlin}. Nevertheless, compelling pieces of evidence show that dimensionality measures are effective determinants of dynamics and activity in networked complex systems. 
The simplest example is provided by networks with the small-world property \cite{Watts1998,Barabasi1999,RPS2001}, which exhibit diameters that grow only logarithmically with the network size $N$ and, consequently, with diverging Hausdorff dimension. We recall that the Hausdorff dimension $d_\mathrm{H}$, also called ``topological dimension'' in the literature \cite{Munoz2010,Moretti2013}, can be computed easily starting from the number $u_i(r)$ of nodes within distance $r$ from node $i$: if $\langle u_i(r)\rangle\sim r^d$ (where $\langle .\rangle$ stands for the average over all nodes in the network), implying  $d_\mathrm{H}=d$
Diverging $d_\mathrm{H}$ typically implies enhanced transmission, signal propagation and high synchronizability.

 A somewhat more complex example of how dimensionality controls activity patterns in networks is provided by hierarchical-modular networks, as models e.g. for brain connectivity \cite{Meunier2010,Fornito}. It was first pointed out that the hierarchical-modular organization of brain regions results in network models of finite Hausdorff dimension $d_\mathrm{H}$ and, at odds with small-world graph topologies, with intrinsically {\it large} diameters \cite{Gallos2012}. The large-world property resulting from finite $d_\mathrm{H}$ in hierarchical-modular networks, a purely structural feature of the network, has been linked to signatures of anomalous activity patterns in brain network models, including among others: sustained activity \cite{Kaiser2007}, sub-diffusive dynamics \cite{Gallos2007}, localization phenomena and stretched criticality in the form of Griffiths phases \cite{Moretti2013,Odor2014,Odor2019}, broad avalanche distributions \cite{Friedman2013,Moretti2013}, states of localized and ``frustrated'' synchronization \cite{Villegas2014, Villegas2016,Millan2018,Odor-Kuramoto,Donetti-optimal}, rounding of first-order phase transitions \cite{Villa2015}, and ergodicity breakdown \cite{Tavani2016,Agliari2016}.  Importantly, some of these anomalous dynamical traits are, in fact, considered essential to the ability of brain networks and of brain-inspired hierarchical architectures to achieve an optimal balance between segregation and integration \cite{Meunier2010}, allowing them to conduct multiple tasks simultaneously, entailing optimal computational capabilities \cite{Agliari2015_PRL}.  Let us also note that a significant part of the above-mentioned phenomenology uses concepts, such as Griffiths phases, first introduced to study (Anderson) localization phenomena in quantum systems described, e.g., by a random tight binding Hamiltonian \cite{Aizenman2015random}, and later extended, for instance, to the Laplacian matrix of a graph \cite{KirschPercolation,Khorunzhiy2006}.

 In this paper, we aim at providing a theoretical foundation for the phenomenological observations of anomalous behavior and localization effects obtained so far in hierarchical-modular networks, establishing for the first time a clear link between the emergence of such localization patterns of activity and synchrony and and spectral properties of the underlying graphs.

 The fundamental concept, allowing us to develop our approach is the spectral dimension $d_\mathrm{s}$ of a graph --as well as an important extension of it, that we call Lifshitz dimension-- which can be defined and measured by simple random walk (RW) analyses \cite{Burioni1995,Burioni2000,Burioni2005}. Given the probability $P_{ij}(t)$ that a random walker starting at time $t_0=0$ from node $i$ arrives at node $j$ after $t$ steps, one can compute the average return probability as $R(t)=\sum_{i=1}^{N}P_{ii}(t)/N$.  If a real positive $d_\mathrm{s}$ exists, such that $R(t)\sim t^{ -\frac{d_\mathrm{s}}{2} }$, $d_\mathrm{s}$ is defined as the (average) spectral dimension of the network \cite{Burioni2005}. While for infinitely large networks further complications arise due to the possibility of transient random walks \cite{Burioni2000}, here we focus on networks of finite, albeit very large size $N$, so that the above definition is intended to hold asymptotically, in the limit of large $N$ and large $t$. The spectral dimension, not unlike the Hausdorff dimension introduced above, is a generalization of the concept of dimension. Actually, in discrete lattices $d_\mathrm{s}=d_\mathrm{H}$, so that both generalizations agree with the Euclidean dimension of the embedding continuum space. This equivalence does not hold in general in heterogeneous networks, nor does it in deterministic fractals \cite{Alexander1982}. We notice in particular that while $d_\mathrm{H}$ is a purely structural measure, $d_\mathrm{s}$ is an observable of a diffusion process operating on the network, and as such it provides us with a probing tool for dynamical signatures of localization and slowing down, and a first approximation in cases, like that of brain activity, with much more complex dynamics.  The relationship between different definitions of a network dimension and their use to predict the emergence of anomalous dynamical patterns thus remains an open question to be fully clarified. For example, it was initially conjectured that Griffiths phases --characterized by string localization features-- in heterogeneous networks can only occur in the finite $d_\mathrm{H}$ case \cite{Munoz2010,Moretti2013}; however, this view was challenged by Mill\'an {\it et al.}, who found that even networks with infinite $d_\mathrm{H}$ can exhibit similar dynamical regimes, provided that $d_\mathrm{s}$ is finite instead \cite{Millan2019}.

With these considerations in mind, we analyse the spectral dimension of hierarchical-modular network models of brain connectivity \cite{Moretti2013}, with the objective of quantifying how the basic traits of brain activity localization, which are captured by these simple network models, are reflected by $d_\mathrm{s}$ and stochastic diffusion null models. To this end, we conducted very-large-scale RW simulations in hierarchical-modular network models of tunable Hausdorff dimension $d_\mathrm{H}$ and computed the average return probabilities $R(t)$.

We chose to work with the  model proposed in \cite{Moretti2013,Safari2017}
for the generation of  synthetic hierarchical-modular networks; this model comes
with a single effective parameter $\alpha$ (the connectivity strength) and an additional parameter $s$ fixing the number of hierarchical levels. In this model the average network degree $\langle k\rangle$ and (asymptotically) the Hausdorff dimension $d_\mathrm{H}$ are both proportional to $\alpha$ so that sparser networks have smaller Hausdorff dimension. While other models proposed in the literature might differ in the choice of parameters \cite{Kaiser2007,Kaiser2010,Friedman2013,Odor2015}, we believe that the conclusions of the current work remain unchanged. In order to capture the large network size limit of the system, we performed random-walk computer simulations on networks of sizes up to $2^{25} \approx 3 \times 10^7$, and for time windows large enough as to ensure that all walkers return to the starting node (which, we recall, is possible because $N$ is finite in our case). 

\begin{figure}
\includegraphics[width=0.45\textwidth]{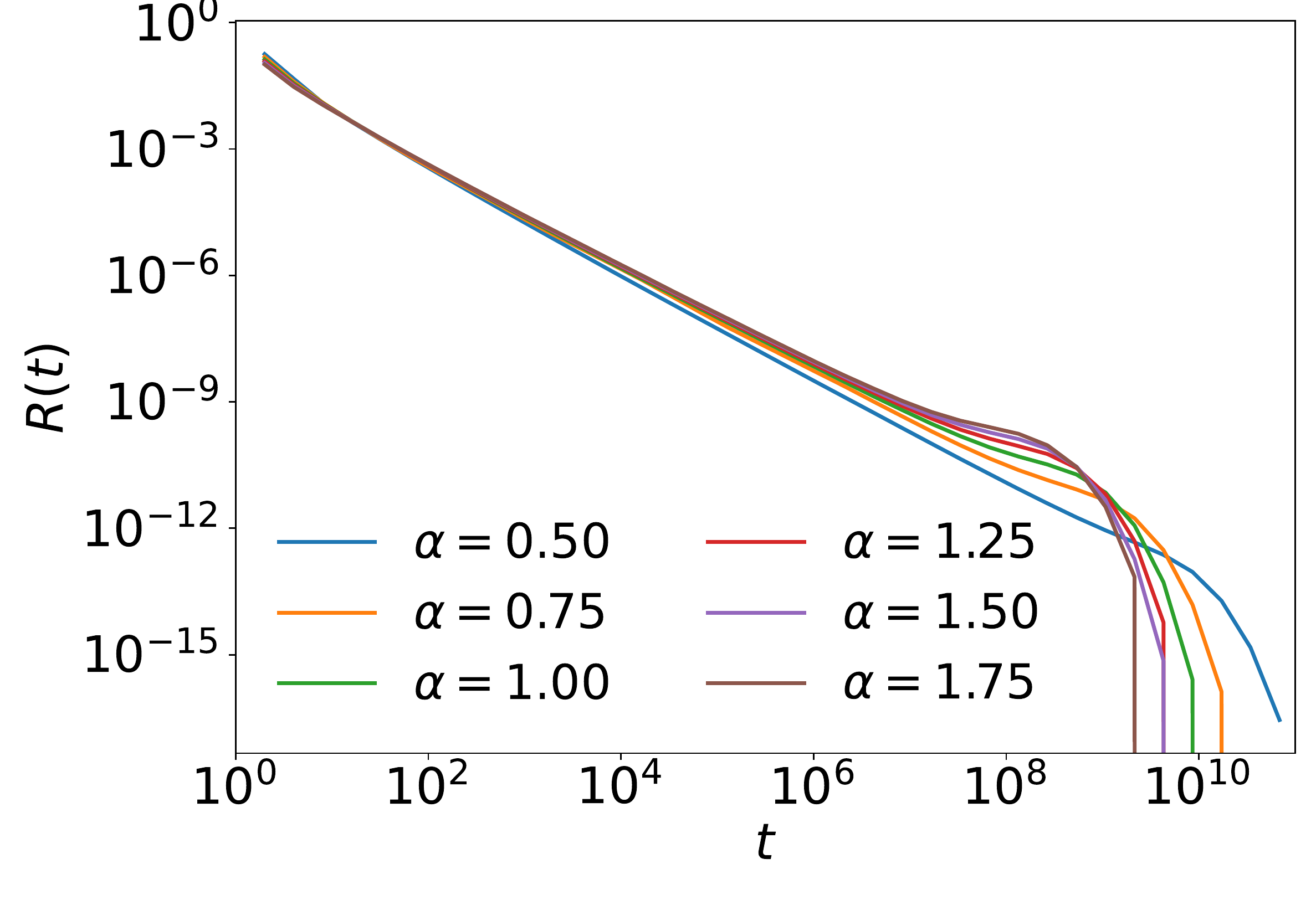}
\caption{Return probabilities for hierarchical-modular networks with $N=2^{25}$, $s=23$ and increasing $\alpha$. Even at large sizes, no clear power law decay is visible and the asymptotic behavior is dominated by a slower (stretched-exponential) tail, before the finite size cut-off takes over. Similar results have been found for smaller sizes and different choices of $s$. 
\label{fig:returnp}}
\end{figure}

Fig. \ref{fig:returnp} shows the return probabilities for a typical choice of parameters ($N=2^{25}$, $s=24$), and for increasing values of $\alpha$ and, thus, increasing Hausdorff dimension $d_\mathrm{H}$. One can immediately see the anomaly in the asymptotic behavior of $R(t)$: while for intermediate n$t$ the curves develop a heavy tail, resembling a power law, the slope of such tails apparently decreases in absolute value upon increasing a$\alpha$, and later develops a non trivial large-$t$ bump, significantly before the finite size cutoff appears. In cases in which the spectral dimension is defined, we would expect the slope to increase in absolute value with $\alpha$, implying that $d_\mathrm{s}$ increases with $d_\mathrm{H}$ and one would expect that behavior to be the asymptotic ($t\to\infty$) one. In the present study, instead, one is forced to conclude that dynamical slowing down is so radical that the asymptotics are given by the excess returns at very large $t$ (the bump in the curves above) and the average spectral dimension is, as a consequence, undefined.      

\begin{figure}
\hspace{-0.5cm}\includegraphics[width=0.5\textwidth]{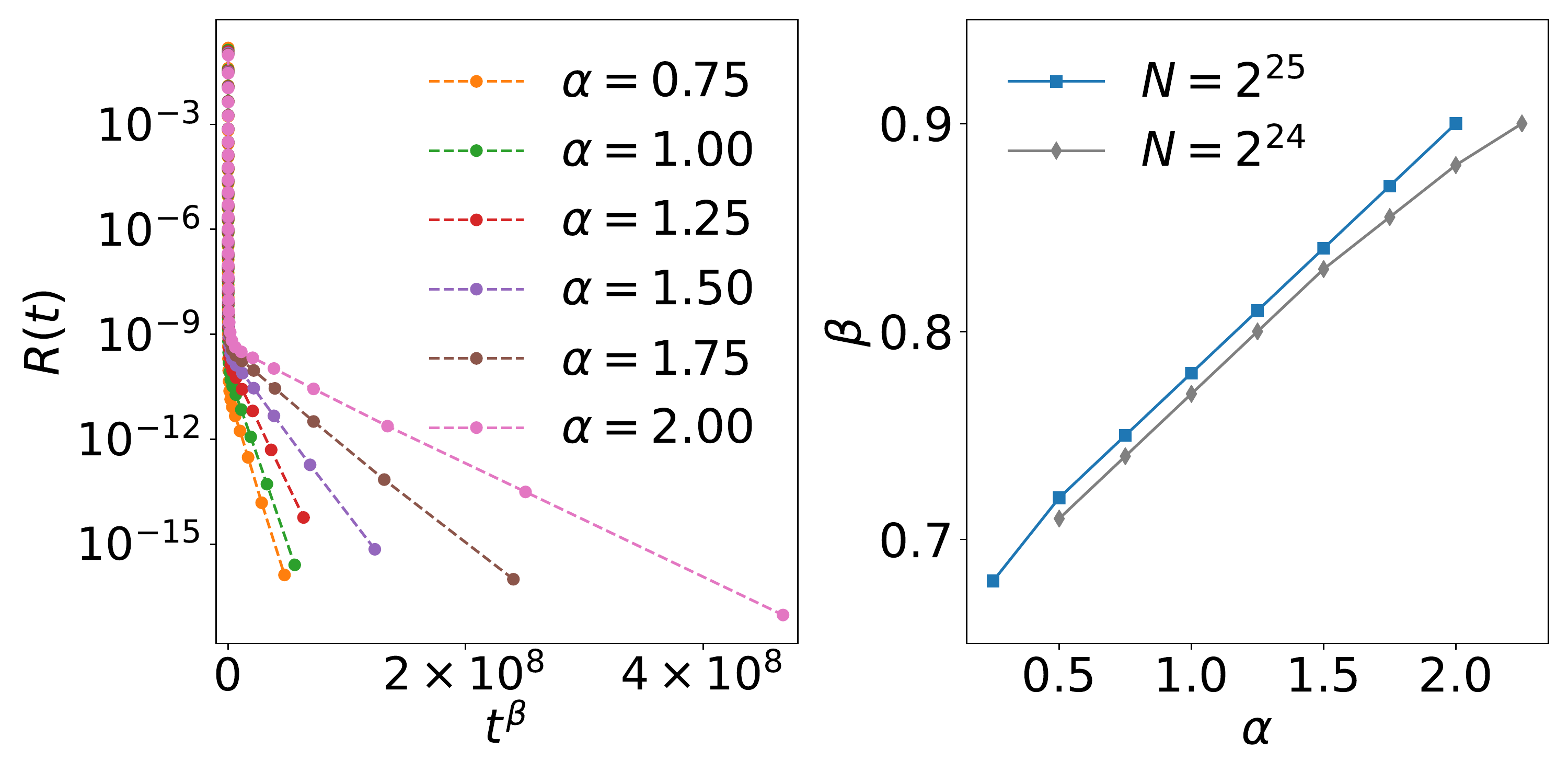}
\caption{Quantitative analysis of the stretched exponential behavior found in Fig. \ref{fig:returnp}. Left: semi-logarithmic-scale plot of the return probability. For each curve (each value of $\alpha$) the value of $\beta$ is chosen, which makes the stretched exponential tail appear as a straight line. Color scheme is as in Fig. \ref{fig:returnp}. Right: the values of $\beta$ used on the left panel ($N=2^{25}$, $s=23$, blue line) as well as for a smaller system ($N=2^{24}$, $s=22$, gray line). 
\label{fig:beta}}
\end{figure}

The absence of a well-defined spectral dimension is an already interesting result within the study localization and dynamical slowing down in models of brain connectivity.
Suppression of diffusion and free flow are considered signatures of anomalously slow dynamical regimes, which ensure the balance between global integration and functional modularity of brain activity \cite{Song2006,Gallos2007,Gallos2012}. To proceed, let us first elucidate the nature of the asymptotic behavior of the return probabilities. Fig. \ref{fig:beta} reveals that the large $t$ dependence of $R(t)$ is dominated by a stretched exponential behavior $R(t) \sim \mathrm{e}^{ - t^\beta}$, governed by a non trivial positive $\beta<1$ (for ease of notation, we measure $t$ in dimensionless units). We call $\beta$ the anomalous exponent, as its value quantifies the dynamical slowing down with respect to the standard scenario where a spectral dimension is defined.
To confirm this view, Fig. \ref{fig:beta} also shows the dependence of $\beta$ on $\alpha$ and thus on the Hausdorff dimension of the network. Sparser networks exhibit stronger anomalous behavior, with the anomaly exponent loosely proportional to the Hausdorff dimension.


In order to make analytical progress, we exploit well-known methods of spectral graph theory \cite{Chung,Mieghem}. Let us write the exact master equation, describing a random walk as a time-continuous Markov process on a generic undirected and unweighted network, encoded in an adjacency matrix $\mathbf{A}$ as follows: \begin{equation}\label{eq:master} \dot{\mathbf{q}}(t) =- \mathbf{L}^\mathrm{RW}\,\mathbf{q}(t), \end{equation} where $\mathbf{q}(t)$ is the column vector, whose generic element $q_i(t)$ represents the probability of the random walker to reach node $i$ at time $t$, and $\mathbf{L}^\mathrm{RW}$ is the random walk Laplacian matrix with elements $L_{ij}^\mathrm{RW}=\delta_{ij} - A_{ij}/k_j$ with $i,j \in 1,2,...N$. Here $A_{ij}$ is the generic element of $\mathbf{A}$, equal to $1$ if nodes $i$ and $j$ are linked and $0$ otherwise, and $k_j=\sum_{i=1}^N A_{ij}$ is the degree of node $j$. In order to compute the solution to Eq.\ref{eq:master}, one can introduce the normalized Laplacian $\mathbf{L}$, defined by the similarity transformation $\mathbf{L}^\mathrm{RW} = \mathbf{D}^\frac{1}{2}\mathbf{L}\mathbf{D}^{-\frac{1}{2}}$, where $\mathbf{D}$ is the (diagonal) degree matrix of generic element $D_{ij}=\delta_{ij}k_j$. $\mathbf{L}$ is symmetric and diagonalizable, and by virtue of their similarity, $\mathbf{L}$ and $\mathbf{L}^\mathrm{RW}$ have the same spectrum of eigenvalues, albeit with different eigenvectors \cite{Chung}. The solutions of Eq.\ref{eq:master} can then be written through the eigen-decomposition of $\mathbf{L}$, as $q_i(t)= \sum_{j=1}^{N}k_i^{1/2} K_{ij}(t-t_0) k_j^{-1/2} q_j(t_0)$, in which we introduced the {\it heat kernel} $\mathbf{K}(t)$ of generic element \cite{heat-kernels}: \begin{equation}\label{eq:heatkernel} K_{ij}(t) = \sum_{m=1}^{N}\mathrm{e}^{-\lambda_m t} V_{im} V_{jm}, \end{equation} where $\lambda_m$ is the $m$-th eigenvalue of $\mathbf{L}$, $V_{im}$ is the $i$-th component of the eigenvector of $\mathbf{L}$ associated with $\lambda_m$, and $t_0=0$ without loss of generality.  This well-known identity allows us to connect the spectral perspective with random-walk simulation results: one can easily see that the average return probability $R(t)$ is related to the trace of $\mathbf{K}(t)$ (or {\it heat trace}) through the simple relationship \cite{heat-kernels}
\begin{equation}\label{eq:return_discrete} R(t)=\frac{1}{N}\sum_{i=1}^{N}K_{ii}(t)=\frac{1}{N}\sum_{m=1}^{N} \mathrm{e}^{-\lambda_m t}.  \end{equation} 
The eigenvalues $\lambda_m$ are all real and non-negative and, under the assumptions that the network is undirected and connected, the $0$ eigenvalue is unique and one can always choose the labeling $0=\lambda_1<\lambda_2\le \lambda_3 \le \cdots$ \cite{Chung}. As a consequence, a random walk always reaches a steady state above a time scale given by the smallest nonzero eigenvalues \cite{Chung}. By making a continuum spectrum approximation for $\lambda\ge \lambda_2$, one can introduce the density of states (eigenvalue density distribution) $\rho(\lambda)$, so that Eq.\ref{eq:return_discrete} can be approximated by its continuum limit \begin{equation}\label{eq:return_continuum}
  R(t) \approx \int \rho(\lambda) \mathrm{e}^{-\lambda t} d\lambda,
\end{equation}
where the integral is dominated by the contribution of the lower spectral edge. Thus, under the present assumptions, the density of states $\rho(\lambda)$ and the return probability are related through a simple Laplace-transform operation. In lattices, which are endowed with an integer spectral dimension, this result is well known and leads to $\rho(\lambda)\sim \lambda^{(d_\mathrm{s}/2) -1}$, which shows the relationship between the Laplacian spectra and the spectral dimension \cite{Burioni2005}. This result is also well known in terms of vibrational frequencies $\omega \propto \sqrt{\lambda}$ in lattices and deterministic fractals, leading to a power-law density of states $\tilde{\rho}(\omega)\sim \omega^{d_\mathrm{s}-1}$ \cite{Alexander1982}.

While the approximation in Eq.\ref{eq:return_continuum} holds in many cases, we expect the conclusions regarding $d_{\mathrm{s}}$ to be radically different in case of hierarchical-modular networks, for which our numerical results reveal anomalous stretched-exponential tails of the return-probability function $R(t)$. This anomaly is indeed reflected in the lower spectral edges and in particular in $\rho(\lambda)$ as we show below. It was hypothesized in the past \cite{Moretti2013} that low eigenvalues of $\mathbf{L}$ in such networks form a continuous spectral tail, a Lifshitz tail, in analogy with the random Hamiltonian operators in a tight-binding Schr\"odinger equation
\cite{Khorunzhiy2006}. In particular, it was noted that Lifshitz tails may be relevant to assess the subcritical dynamics in models of epidemic spreading, in the framework of a linearized quenched mean-field approximation \cite{Odor2014}.
The random-walk problem that we study here, instead, can always be mapped exactly to the quantum problem, with the energy eigenvalues $E_m$ of the quantum problem being replaced by the Laplacian eigenvalues $\lambda_m$ \cite{Khorunzhiy2006}. Under the hypothesis of Lifshitz tails, the integrated density of states of $\mathbf{L}$ (i.e. the cumulative eigenvalue density distribution) is expected to exhibit a tail of the general form \cite{Aizenman2015random} 
\begin{equation}\label{eq:lifshitzids}
\mathcal{N}(\lambda) = c_1 \mathrm{exp}\left[c_2 \left(\lambda -\lambda_0\right)^{-\frac{d_\mathrm{L}}{2}} \right],
\end{equation}
where $\lambda_0$ is the lower bound of the continuum spectrum, which we can set equal to $0$ in the case of hierarchical-modular networks, as they possess vanishing spectral gaps ($0<\lambda_2 \ll 1$) \cite{Moretti2013}, and $c_1$ and $c_2$ are constants. The real number $d_\mathrm{L}$ coincides with the space Euclidean dimension in the original Lifshitz argument for  continuum quantum problems. In the present discrete classical case, since that we cannot yet provide a Lifshitz-like argument, we simply name $d_\mathrm{L}$ the \emph{Lifshitz dimension} of the problem.
In the light of the above considerations, the density of states $\rho(\lambda)=d\mathcal{N}/d\lambda$ is dominated by the following low-$\lambda$ tail
\begin{equation}\label{eq:lifshitzdos}
\rho(\lambda) \sim \mathrm{exp}(c_2 \lambda ^{-\frac{d_\mathrm{L}}{2}} ).   
\end{equation}
Observe that Eq.\ref{eq:lifshitzdos} differs significantly from the above power-law relationship for lattices $\rho(\lambda)\sim \lambda^{(d_\mathrm{s}/2) -1}$. Is such a difference a spectral signature of the anomalous behaviour (i.e.  the stretched-exponential tail of the return probabilities $R(t)$ and the lack of a well-defined spectral dimension $d_\mathrm{s}$) encountered in RW simulations? As the two representations connect through Eq.\ref{eq:return_continuum}, one can compute $R(t)$ as in Eq.\ref{eq:return_continuum}, using the hypothesis of a Lifshitz tail from Eq.\ref{eq:lifshitzdos}. While the integral involved in the calculation, of the form $w(t)=\int \mathrm{e}^{g(\lambda,t)} d\lambda$ with $g(\lambda,t)=c_2 \lambda^{-d_\mathrm{L}/2}-\lambda t$,  is highly non-trivial in the case of real $d_\mathrm{L}$, here we are only interested in its asymptotic $t\to\infty$ behavior, which is captured by the values of $\lambda$ for which $g(\lambda,t)$ is maximum. This is readily obtained through the saddle-point approximation $w(t)\approx \mathrm{e}^{g(\lambda^*,t)}$, with $\lambda^*$ the location of such maximum, leading to the final result
\begin{equation}\label{eq:return_lifshitz}
R(t)\sim \exp\left(-t^{\frac{d_\mathrm{L}}{2+d_\mathrm{L}}}\right).
\end{equation}
Eq.\ref{eq:return_lifshitz} remarkably recovers a stretched exponential tail behavior, as reported above for computer simulations, confirming for the first time that the dynamical slowing and lack of spectral dimension can be attributed to the existence of Lifshitz tails and providing us with an interpretation of the anomalous exponent in terms of the Lifshitz dimension, $d_\mathrm{L}$
\begin{equation}\label{eq:beta}
\beta=\frac{d_\mathrm{L}}{2+d_\mathrm{L}}.
\end{equation}
We can conclude that in the present hierarchical-modular network model, not only Lifshitz tails explain the anomalous dynamics and the lack of a well-defined spectral dimension, but also $d_\mathrm{L}$ provides us with a meaningful dimensionality measure, generalizing the behavior of quantum systems in the continuum, where the Lifshitz dimension identifies the spatial dimension.

\begin{figure}
\hspace{-0.5cm}\includegraphics[width=0.4\textwidth]{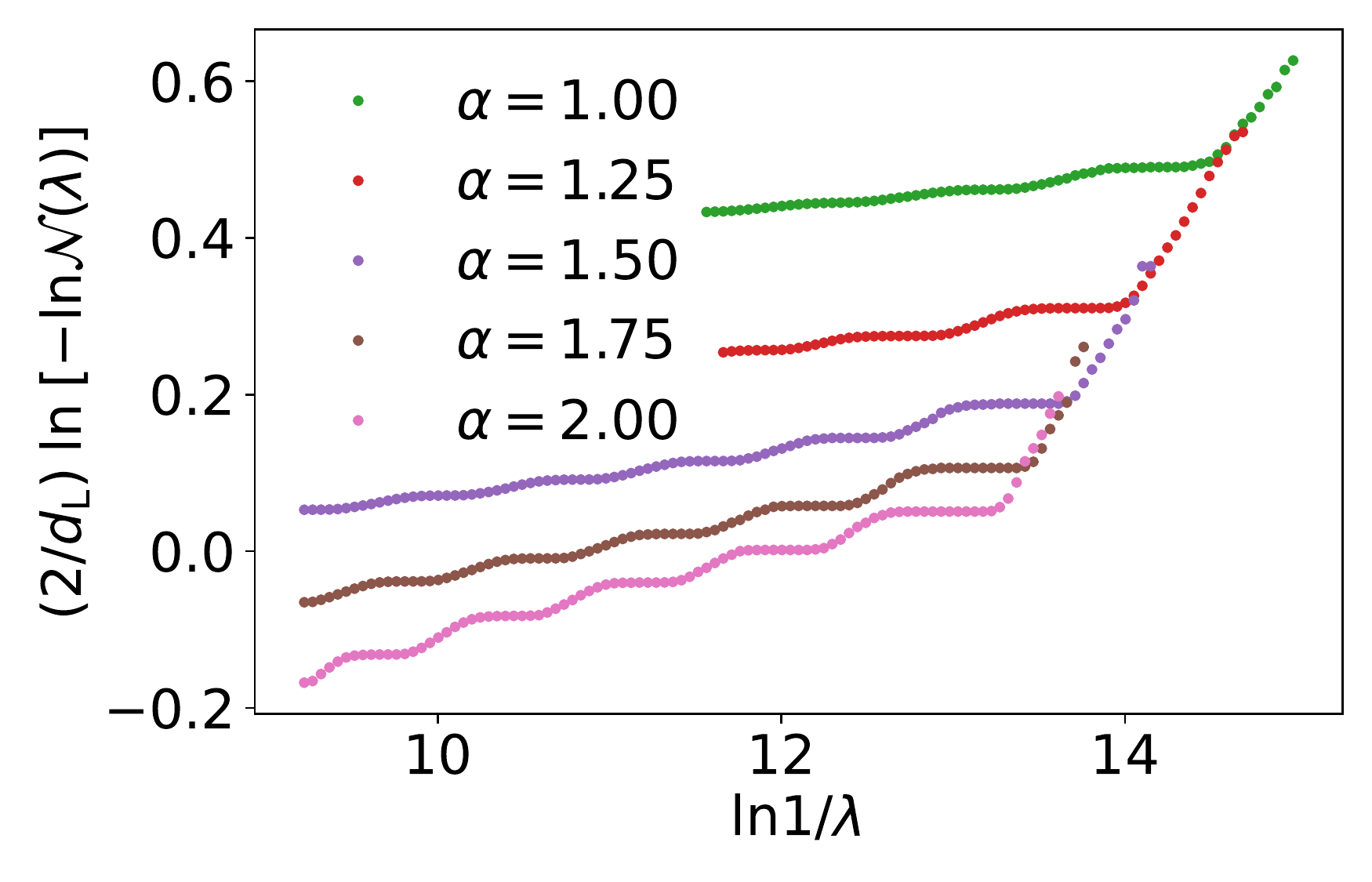}
\caption{Lower spectral edge of $\mathbf{L}$. Using the rescaling from Eq.\ref{eq:tailscaling}, Lifshitz tails appear as straight lines. By choosing values of $d_\mathrm{L}$ obtained from the RW simulation results through Eq.\ref{eq:beta}, Lifshitz tails collapse in a single curve, confirming that the anomalous dynamical behavior has its origin in the spectral properties of $\mathbf{L}$.}
\label{fig:lifshitz}
\end{figure}

So far, we have only hypothesized that the lower spectral edge of $\mathbf{L}$ exhibits a Lifshitz tail of the form given by Eqs.\ref{eq:lifshitzids} and \ref{eq:lifshitzdos}. Now we corroborate such a hypothesis by verifying, not only that the integrated density of states $\mathcal{N}(\lambda)$ obeys the tail behavior in Eq.\ref{eq:lifshitzids}, but also that the exponent $d_\mathrm{L}/2$ governing it generates the anomalous exponent $\beta$ of the dynamical simulation, as predicted by the result in Eq.\ref{eq:beta}. To this end, we notice that according to the prediction above, one expects 
\begin{equation}\label{eq:tailscaling}
\ln[-\ln\mathcal{N}(\lambda)]\sim (d_\mathrm{L}/2)\ln 1/\lambda
\end{equation}
for large $1/\lambda$. We obtain $d_\mathrm{L}=2\beta/(1-\beta)$ from Eq.\ref{eq:beta}, and using the values of $\beta$ obtained from the initial random walk simulations one can easily verify Eq.\ref{eq:tailscaling} by computing the lower spectral edges of hierarchical-modular networks of the same type. Computational results, shown in Fig. \ref{fig:lifshitz} clearly confirm the linear dependence predicted
by Eq.\ref{eq:tailscaling} for small values of $\lambda$. In other words, 
the prediction based on the Lifshitz tails assumption is correct: the spectra of hierarchical-modular networks exhibit Lifshitz tails, with an associated Lifshitz dimension $d_\mathrm{L}$, and their concomitant anomalous dynamical behavior is controlled by  $d_\mathrm{L}$.

Moreover, even if not explicitly analyzed here, the eigenvalues in the Lifshitz tail have strongly localized eigenvectors, meaning that
  their components vanish almost everywhere except in specific network locations such as moduli \cite{Chung,Moretti2013}. This property, known as {\it eigenvector localization}  \cite{Aizenman2015random}  is analogous to the case of disordered quantum systems, where localization stands for absence of diffusion, 
and has been observed in models of epidemic spreading on networks \cite{Goltsev2012,Odor2014,PastorSatorras2016,PastorSatorras2018}, as well in problems inspired by brain connectivity \cite{Moretti2013,Odor2015}  and biological materials \cite{Moretti2019}.  In the particular case of brain dynamics, localization can play a key role in allowing for task segregation.

It is noteworthy that the emergence of {\it classical} Lifshitz tails in network spectra has been rigorously proved in Erd\H{o}s R\'enyi graphs below the percolation threshold \cite{Khorunzhiy2006}, i.e. for networks that have not yet developed a giant connected component. Our results suggest a remarkable property of hierarchical-modular networks, which exhibit Lifshitz tails {\it while} being connected, i.e. while possessing a single connected component. We believe that Lifshitz tails, and the resulting anomalous dynamical behavior, may be observed in general in network models exhibiting similar localization properties, such as hierarchical trees displaying {\it patchy} percolation  \cite{Boettcher2009}, and dense hierarchical Dyson networks \cite{Agliari2015_PRL}, where ergodicity is known to  break down in the thermodynamic limit \cite{Tavani2016,Agliari2016}. 
In fact, we propose the emergence of Lifshitz tails and their associated Lifshitz dimension as a criterion for the existence of localization in networks.

Our focus on the spectral dimension and its undefined nature in hierarchical networks allows us to establish a connection between network structural properties and anomalous dynamics in systems such as brain networks, where the ongoing structure vs. function debate has long dealt with the issues of relating activity patters to specific anatomical arrangements, or alternatively presenting them as {\it emergent}, or {\it self-organized}. Our results clearly show a connection between dynamical slowing-down, localization properties, and Laplacian spectra; let us emphasize that such a correspondence is clean-cut because of the simplicity of the random walk model, and possibly because of the simplifying assumptions in the choice of our network model.  While diffusion --lacking any form of non-linearity-- is arguably a very crude simplification of neural dynamics on the structurally complex human connectome \cite{Fornito}, it has been found that the eigenvectors of a diffusion problem on the connectome are relevant in predicting functional patterns of neural activity \cite{Atasoy1,Atasoy2}. More in general, the Laplacian matrix provides the linearization of oscillator models near a synchronization transition \cite{Arenas2006} and the relevance of its spectrum in the problem of brain synchronization has been discussed in the literature and related to the observation of frustrated synchronization \cite{Villegas2014, Villegas2016,Millan2018,Millan2019}. While some of those results were based on the hypothesis of the existence of Lifshitz tails, here we are able to prove such a hypothesis and rationalize those results within a proper theoretical framework, where the slowing down of synchronization processes is governed by the Lifshitz dimension $d_\mathrm{L}$, which effectively tunes the dynamical anomalies. Let us also note that, while the approach here resorts to unweighted networks, the Laplacian formalism lends itself to the introduction of weights \cite{Tavani2016,Agliari2016}, a quantity that in neuroimaging encodes the number of connections between pairs of brain regions. Beyond the case of pairwise interactions,  recent advances in integrating the concepts of diffusion and spectral dimension within the broader field of algebraic topology and simplicial complexes \cite{Torres2020Complexity,Bianconi2020JSTAT,Reitz2020} provide a promising avenue to strengthen the theoretical framework for the localization phenomena that we discuss here to describe, for instance, systems with higher-order interactions between their components/nodes.

In conclusion, we have established a theoretical framework for the prediction of anomalous dynamics in hierarchical network models of interest in brain modelling.  To our knowledge, hierarchical-modular networks constitute the first heterogeneous network model displaying Lifshitz tails above the percolation threshold, and the first not to exhibit a power-law behavior for the average return probability and a well-defined spectral dimension. Being able to connect these two singular features allows us to rationalize previous experimental observations of activity localization in the brain and their numerical models, where spectral anomalies and Lifshitz tails where only hypothesized. We believe that these results will stimulate interest and further work, in e.g. computational neuroscience, as a way to advance the knowledge on how the brain achieves an optimal balance between segregation (localization on specific moduli) and integration.  In particular,  we plan to extend our approach to novel network models of brain connectivity, including important architectural features such
  as  a prominent core-periphery or rich-club organization \cite{Rich-club,Core-periphery,Asymmetry}.
Finally, we are confident that the present framework will provide us with more powerful tools for the tunability and controllability of network models exhibiting strong localization, relevant in the design of synthetic networks for brain-inspired neuromorphic computing.

\begin{acknowledgments}
We acknowledge the Deutsches Forschungsgemeinschaft through grants \mbox{MO 3049/1-1}, \mbox{MO 3049/3-1} and GRK 2423 FRASCAL, the Spanish Ministry and Agencia Estatal de investigaci{\'o}n (AEI) through grants \mbox{FIS2017-84256-P} (European
  Regional Development Fund(ERDF), as well as the Consejer{\'\i}a de
  Conocimiento, Investigaci{\'o}n y Universidad, Junta de
  Andaluc{\'\i}a and ERDF, Ref.  \mbox{A-FQM-175-UGR18} and \mbox{SOMM17/6105/UGR} and for financial support.
\end{acknowledgments}

\input{submit.bbl}

\end{document}

%% file: submit.bbl
%